\title{NARROW ESCAPE, part II: The circular disk}
\author{A. Singer \thanks{Department of Applied Mathematics,
Tel-Aviv University, Ramat-Aviv, 69978 Tel-Aviv, Israel, e-mail:
amits@post.tau.ac.il}\,,\ \ Z. Schuss\thanks{Department of
Mathematics, Tel-Aviv University, Tel-Aviv 69978,  Israel, e-mail:
schuss@post.tau.ac.il.}\,,\ \ D. Holcman\thanks{Department of
Mathematics, Weizmann Institute of Science, Rehovot 76100 Israel,
e-mail holcman@wisdom.weizmann.ac.il. }
\thanks{Keck Center, department of Physiology, UCSF, 513 Parnassus
Ave, San Francisco 94143 USA, e-mail holcman@phy.ucsf.edu.}
}
\documentclass[12pt]{article}
\usepackage{float}
\usepackage{amsmath}
\usepackage{epsfig, graphicx} 
\newcommand{\mb}[1]{\mbox{\boldmath$#1$}}
\newcommand{\p}{\partial}
\newcommand{\ds}{\displaystyle}
\newcommand{\beq}{\begin{eqnarray}}
\newcommand{\beqq}{\begin{eqnarray*}}
\newcommand{\eeq}{\end{eqnarray}}
\newcommand{\eeqq}{\end{eqnarray*}}
\newcommand{\x}{\mbox{\boldmath$x$}}

\def\ds#1{\displaystyle{#1}}

\begin{document}
\numberwithin{equation}{section} \maketitle

\begin{abstract}
 We consider Brownian motion in a circular disk $\Omega$, whose
boundary $\p\Omega$ is reflecting, except for a small arc,
$\p\Omega_a$, which is absorbing. As $\varepsilon=|\partial
\Omega_a|/|\partial \Omega|$ decreases to zero the mean time to
absorption in $\p\Omega_a$, denoted $E\tau$, becomes infinite. The
narrow escape problem is to find an asymptotic expansion of
$E\tau$ for $\varepsilon\ll1$. We find the first two terms in the
expansion and an estimate of the error. The results are extended
in a straightforward manner to planar domains and two-dimensional
Riemannian manifolds that can be mapped conformally onto the disk.
Our results improve the previously derived expansion for a general
smooth domain, $E\tau = \ds{
\frac{|\Omega|}{D\pi}}\left[\log\ds{\frac{1}{\varepsilon}}+O(1)\right],$
($D$ is the diffusion coefficient) in the case of a circular disk.
We find that the mean first passage time from the center of the
disk is $E[\tau\,|\,
\x(0)=\mb{0}]=\ds{\frac{R^2}{D}}\left[\log\ds{\frac{1}{\varepsilon}}
+ \log 2 +\ds{ \frac{1}{4}} + O(\varepsilon)\right]$. The second
term in the expansion is needed in real life applications, such as
trafficking of receptors on neuronal spines, because
$\log\ds{\frac{1}{\varepsilon}}$ is not necessarily large, even
when $\varepsilon$ is small. We also find the singular behavior of
the probability flux profile into $\p\Omega_a$ at the endpoints of
$\p\Omega_a$, and find the value of the flux near the center of
the window.
\end{abstract} \maketitle

\section{Introduction}
The expected lifetime of a Brownian motion in a bounded domain,
whose boundary is reflecting, except for a small absorbing
portion, increases indefinitely as the absorbing part shrinks to
zero. The narrow escape problem is to find an asymptotic expansion
of the expected lifetime of the Brownian motion in this limit. The
narrow escape problem in three dimensions has been studied in the
first paper of this series \cite{NarrowEscape1}, where is was
converted to a mixed Dirichlet-Neumann boundary value problem for
the Poisson equation in the domain. This is a well known problem
of classical electrostatics (e.g., the electrified disk problem
\cite{Jackson}), elasticity (punch problems), diffusion and
conductance theory, hydrodynamics, and acoustics
\cite{Sneddon}-\cite{Vinogradov}. It dates back to Helmholtz
\cite{Helmholtz} and Lord Rayleigh \cite{Rayleigh} and has been
extensively studied in the literature for special geometries.

The study of the two-dimensional narrow escape problem began in
\cite{Holcman} in the context of receptor trafficking on
biological membranes \cite{Choquet}, where a leading order
expansion of the expected lifetime was constructed for a general
smooth planar domain. In this paper we present a thorough analysis
of the narrow escape problem for the circular disk and note that
our calculations apply in a straightforward manner to any simply
connected domain in the plane that can be mapped conformally onto
the disk. According to Riemann's mapping theorem
\cite{Markushevich}, this covers all simply connected planar
domains whose boundary contains at least one point. The same
conclusion holds for the narrow escape problem on two-dimensional
Riemannian manifolds that are conformally equivalent to a circular
disk. The biological problem of receptor trafficking on membranes
is locally planar, but globally it is a problem on a Riemannian
manifold. The narrow escape problem of non-smooth domains that
contain corners or cusp points at their boundary is treated in the
third part of this series \cite{NarrowEscape3}, where the
conformal mapping method is demonstrated.

The specific mathematical problem can be formulated as follows. A
Brownian particle diffuses freely in a disk $\Omega$, whose
boundary $\p\Omega$ is reflecting, except for a small absorbing
arc $\p\Omega_a$. The ratio between the arclength of the absorbing
boundary and the arclength of the entire boundary is a small
parameter
 $$\varepsilon = \ds{\frac{|\partial
 \Omega_a|}{|\partial \Omega|}} \ll 1.$$
The mean first passage time to $\p\Omega_a$, denoted $E\tau$,
becomes infinite as $\varepsilon\to0$. The asymptotic expansion of
$E\tau$ for $\varepsilon\ll1$ was considered for the particular
case when $\p\Omega_a$ is a disjoint component of $\p\Omega$ in
\cite[and references therein]{Pinsky}. This case differs from the
case at hand in that the absorption probability flux density in
the former is regular, while in the latter it is singular. It was
shown in \cite{Holcman} that $E\tau$ for the narrow escape problem
in a general planar domain $\Omega$ has the asymptotic form
\begin{equation}
\label{eq:holcman} E\tau = \ds{
\frac{|\Omega|}{D\pi}}\left[\log\frac{1}{\varepsilon}+O(1)\right],
\end{equation}
where $|\Omega|$ is the area of $\Omega$, and $D$ is the diffusion
coefficient. This leading order asymptotics has the drawback that
$\log\varepsilon$ can be $O(1)$ when $\varepsilon\ll1$. Thus the
second term in the expansion is needed. For the particular case of
a circular disk an approximate value for the correction was given
in \cite{Holcman}. In contrast, the asymptotics of $E\tau$ for a
three dimensional ball of radius $R$ with an absorbing window of
radius $\varepsilon R$ is \cite{NarrowEscape1}
 $$E\tau=\ds{\frac{|\Omega|}{4D\varepsilon R}}\left[1+O(\varepsilon \log
 \varepsilon)\right],$$
so the leading order term is much larger than the correction term
if $\varepsilon$ is small. The difference in the asymptotic form
of $E\tau$ stems from the different singularities of the Neumann
function in two and three dimensions: it is logarithmic in two
dimensions and has a pole in three dimensions.

Our computations are based on the mixed boundary value techniques
of \cite{Sneddon}. They reveal the singularity of the absorption
flux in the absorbing arc $\p\Omega_a$. Specifically, the
singularity is $(\varepsilon^2-s^2)^{-1/2}$, where $s$ is the
(dimensionless) arclength measured from the center of
$\p\Omega_a$, and attains the values $s=\pm\varepsilon$ at the
endpoints.

The exit time vanishes at the absorbing boundary, and is small
near the absorbing boundary, but it attains large and almost
constant values of order $\log\ds{\frac{1}{\varepsilon}}$ inside
the domain. We show that this ``jump'' occurs in a small boundary
layer of size $O\left(\varepsilon
\log\ds{\frac{1}{\varepsilon}}\right)$. We calculate the average
exit time, where the averaging is against a uniform initial
distribution in the disk, the time to exit from the center, and
the maximum mean exit time, attained at the antipodal point to the
center of the absorbing window.

The mean first passage time (MFPT) from the center of the disk is
\begin{equation}
\label{main:mfpt-center} E[\tau\,|\,\x(0)=\mb{0}]
=\frac{R^2}D\left[\log\frac{1}{\varepsilon} + \log 2 + \frac{1}{4}
+ O(\varepsilon)\right],
\end{equation}
the MFPT, averaged with respect to an initial uniform distribution
in the disk is
\begin{equation}
\label{main:averaged} E\tau =
\frac{R^2}D\left[\log\frac{1}{\varepsilon} + \log 2 + \frac{1}{8}
+ O(\varepsilon)\right],
\end{equation}
and the maximal value of the MFPT is attained on the
circumference, at the antipodal point to the center of the hole,
\begin{equation}
\label{main:v_max} \max_{\x\in\Omega}E[\tau\,|\,\x] =
E[\tau\,|\,r=1, \theta=0] =
\frac{R^2}D\left[\log\frac{1}{\varepsilon} + 2\log2 +
O(\varepsilon)\right].
\end{equation}

{\bf The boundary layer analysis of $E\tau$ can be applied to the
approximation of the first eigenfunction and eigenvalue of the
mixed Neumann-Dirichlet boundary value problem with a small
Dirichlet window on the boundary. This problem arises in the
construction of the first eigenfunction and eigenvalue of the
Neumann problem in a domain that consists of two domains (e.g.,
circular disks) connected by a narrow channel \cite{Berez},
\cite{Weiss}.

Specifically, it is easy to see that
 \beq
 E\tau=\sum_{n=0}^\infty\frac1{\lambda_n}\sim\frac1{\lambda_0}\,\label{ev}
 \eeq
where $0<\lambda_0\ll\lambda_1<\cdots$ are the eigenvalues of the
mixed problem and the MFPT is also averaged with respect to the
initial point. The first eigenfunction $u_0$ of the mixed problem
is differs from the first eigenfunction of the Neumann problem,
which is $v_0=1$, only in a boundary layer about the small window.
Thus $u_0$ is a small perturbation (in $L^2$ norm) of $v_0=1$. It
follow that $u_0/\lambda_0$ differs from $E\tau$ only in the
boundary layer.}

\section{Solution of a mixed boundary value problem}
\label{sec:disk} In non-dimensional variables the narrow escape
problem concerns Brownian motion inside the unit disk, whose
boundary is reflecting but for a small absorbing arc of length
$2\varepsilon$ (see Fig.\ref{f:disk}). In polar coordinates
$\x=(r,\theta)$ the MFPT
 \[v(r,\theta)=E[\tau\,|\,\x(0)=(r,\theta)],\]
is the solution to the mixed Neumann-Dirichlet inhomogeneous
boundary value problem (see, e.g. \cite{Schuss})
\begin{eqnarray}
\Delta v(r,\theta) & = & -1, \quad r<1, \quad\mbox{for}\quad 0\leq
\theta < 2\pi,
\nonumber \\
&&\nonumber\\
v(r,\theta)\bigg|_{r=1} & = & 0, \quad\mbox{for}\quad |\theta-\pi|<\varepsilon, \nonumber \\
&&\nonumber\\
 \frac{\partial v(r,\theta)}{\partial r}\bigg|_{r=1}
& = & 0, \quad\mbox{for}\quad |\theta-\pi| \rangle \varepsilon,
\end{eqnarray}
which is reduced by the substitution
 \beq
u=v- \frac{1-r^2}{4}\label{eq:u-v}
 \eeq
to the mixed Neumann-Dirichlet problem for the Laplace equation
\begin{eqnarray}
\Delta u(r,\theta) & = & 0, \quad \mbox{for}\quad r<1, \quad 0
\leq \theta < 2\pi,\nonumber \\
&&\nonumber\\
 u(r,\theta)\bigg|_{r=1} & = & 0, \quad \mbox{for}\quad|\theta-\pi|
< \varepsilon,\label{eq:f-dirichlet} \\
&&\nonumber\\
 \frac{\partial u(r,\theta)}{\partial r}\bigg|_{r=1}
& = & \frac{1}{2}, \quad\mbox{for}\quad |\theta-\pi|\rangle
\varepsilon. \nonumber
\end{eqnarray}

We adapt the method of \cite{Sneddon} to the solution of
(\ref{eq:f-dirichlet}). Separation of variables suggests that
\begin{equation}
\label{eq:u-separation} u(r,\theta) = \frac{a_0}{2} +
\sum_{n=1}^\infty a_n r^n \cos n\theta,
\end{equation}
where the coefficients $\{a_n\}$ are to be determined by the
boundary conditions
\begin{eqnarray}
\label{eq:boundary-1} u(r,\theta)\bigg|_{r=1} & = & \frac{a_0}{2} +
\sum_{n=1}^\infty
a_n \cos n\theta = 0, \quad \mbox{for}\quad\pi-\varepsilon < \theta \leq \pi, \\
&&\nonumber\\
 \label{eq:boundary-2} \frac{\partial
u(r,\theta)}{\partial r}\bigg|_{r=1} & = & \sum_{n=1}^\infty n a_n
\cos n\theta = \frac{1}{2}, \quad\mbox{for}\quad 0 \leq \theta <
\pi-\varepsilon.
\end{eqnarray}
We identify this problem with problem (5.4.4) in \cite{Sneddon},
where general functions appear on the right hand sides of
equations (\ref{eq:boundary-1}) (\ref{eq:boundary-2}). Due to the
invertibility of Abel's integral operator, the equation
\begin{equation}
\frac{a_0}{2} + \sum_{n=1}^\infty a_n \cos n\theta = \cos
\frac{1}{2}\theta \int_{\theta}^{\pi-\varepsilon}
\frac{h_1(t)\,dt}{\sqrt{\cos\theta - \cos t}},
\end{equation}
defines $h_1(t)$ uniquely for $0 \leq t < \pi-\varepsilon$. The
coefficients are given by
\begin{eqnarray}
a_n & = & \frac{2}{\pi} \int_0^{\pi-\varepsilon} \cos n\theta \cos
\frac{1}{2}\theta \,d\theta \int_{\theta}^{\pi-\varepsilon}
\frac{h_1(t)\,dt}{\sqrt{\cos\theta - \cos t}} \nonumber \\
&&\nonumber\\
 & = & \frac{1}{\pi} \int_0^{\pi-\varepsilon}
h_1(t)\,dt \int_0^t \frac{\cos \left(n+\frac{1}{2}\right)\theta +
\cos \left( n-\frac{1}{2}\right)\theta}{\sqrt{\cos\theta-\cos
t}}\,d\theta.
\end{eqnarray}
The integral
\begin{equation}
P_n(\cos u) = \frac{\sqrt{2}}{\pi} \int_0 ^u \frac{\cos\left(
n+\frac{1}{2}\right)\theta }{\sqrt{\cos \theta-\cos u}}\, d\theta,
\end{equation}
is  Mehler's integral representation of representation of the
Legendre polynomial \cite{Stegun}. It follows that
\begin{equation}
\label{eq:a_n} a_n = \frac{1}{\sqrt{2}} \int_0^{\pi-\varepsilon}
h_1(t) [P_n(\cos t) + P_{n-1}(\cos t)]\,dt,
\end{equation}
for $n> 0$, and
\begin{equation}
\label{a_0} a_0 = \frac{2}{\pi} \int_0 ^{\pi-\varepsilon}
h_1(t)\,dt \int_0^t \frac{\cos \frac{1}{2}\theta}{\sqrt{\cos
\theta - \cos t}}\,d\theta = \sqrt{2} \int_0^{\pi-\varepsilon}
h_1(t)\,dt.
\end{equation}
Integration of (\ref{eq:boundary-2}) gives
\begin{equation}
\sum_{n=1}^\infty a_n \sin n\theta = \frac{1}{2}\theta,
\quad\mbox{for}\quad 0 \leq \theta < \pi-\varepsilon.
\end{equation}
Changing the order of summation and integration yields
\begin{equation}
\int_0^{\pi-\varepsilon} h_1(t) \frac{1}{\sqrt{2}}
\sum_{n=1}^\infty [P_n(\cos t) + P_{n-1}(\cos t)]\sin n\theta \,dt
= \frac{1}{2}\theta.
\end{equation}
Using  \cite[eq.(2.6.31)]{Sneddon},
\begin{equation}
\label{eq:heavyside} \frac{1}{\sqrt{2}}\sum_{n=1}^\infty [P_n(\cos
t) + P_{n-1}(\cos t)]\sin n\theta = \frac{\cos\frac{1}{2}\theta
H(\theta-t)}{\sqrt{\cos t - \cos \theta}},
\end{equation}
we obtain
\begin{equation}
\int_0^{\theta} \frac{h_1(t)\,dt}{\sqrt{\cos t- \cos \theta}} =
\frac{\theta}{2\cos\frac{1}{2}\theta},\quad\mbox{for}\quad0 \leq
\theta < \pi -\varepsilon.\label{Abeltype}
\end{equation}
The solution of the Abel-type integral equation (\ref{Abeltype})
is given by
\begin{equation}
\label{eq:h_1} h_1(t) = \frac{1}{\pi} \frac{d}{dt} \int_0^t
\frac{u \sin \ds{\frac{u}{2}}}{\sqrt{\cos u - \cos t}}\,du.
\end{equation}
Together with (\ref{a_0}) this gives
\begin{equation}
\label{eq:int} a_0 = \frac{\sqrt{2}}{\pi} \int_0^{\pi-\varepsilon}
\frac{u \sin \ds{\frac{u}{2}}}{\sqrt{\cos u + \cos
\varepsilon}}\,du.
\end{equation}

We expect the function $u(r,\theta)$, closely related to the MFPT,
to be almost constant in the disk, except for a boundary layer
near the absorbing arc. The value of this constant is $a_0$,
because all other terms of expansion (\ref{eq:u-separation}) are
oscillatory.

\subsection{Small $\varepsilon$ asymptotics} The results of the previous
section are independent of the value of $\varepsilon$. Here we
find the asymptotic of $a_0$ for $\varepsilon \ll 1$. Substituting
\begin{equation}
\label{eq:magic-substitution} s=\sqrt{\frac{\cos u + \cos
\varepsilon}{2}}
\end{equation}
in the integral (\ref{eq:int}) yields
\begin{eqnarray}
\label{eq:a_0} a_0 & = &
\frac{4}{\pi}\int_0^{\cos(\varepsilon/2)}\frac{\arccos
\sqrt{s^2+\sin^2\ds{\frac{\varepsilon}{2}}}}{\sqrt{s^2+\sin^2
\ds{\frac{\varepsilon}{2}}}}\,ds \nonumber \\
&&\nonumber\\
& = & 2\int_0^{\cos(\varepsilon/2)}\frac{1}{\sqrt{s^2+\sin^2
\frac{\varepsilon}{2}}}\,ds -
\frac{4}{\pi}\int_0^{\cos(\varepsilon/2)}\frac{\arcsin
\sqrt{s^2+\sin^2\ds{\frac{\varepsilon}{2}}}}{\sqrt{s^2+\sin^2
\ds{\frac{\varepsilon}{2}}}}\,ds \nonumber \\
&&\nonumber\\
 & = & 2\log\left(1+\cos\frac{\varepsilon}{2}\right)
- 2\log\sin\frac{\varepsilon}{2} -
\frac{4}{\pi}\int_0^{\cos(\varepsilon/2)}\frac{\arcsin
\sqrt{s^2+\sin^2\ds{\frac{\varepsilon}{2}}}}{\sqrt{s^2+\sin^2
\ds{\frac{\varepsilon}{2}}}}\,ds \nonumber \\
&&\nonumber\\
& = & -2\log \frac{\varepsilon}{2} + 2\log2 - \frac{4}{\pi}
\int_0^1\frac{\arcsin s}{s}\,ds +
O(\varepsilon) \nonumber \\
&&\nonumber\\
& = & -2\log\frac{\varepsilon}{2} + O(\varepsilon),
\end{eqnarray}
because $\ds{\int_0^1 \frac{\arcsin s}{s}}\,ds =
\ds{\frac{\pi}{2}} \log2$. The substitution
(\ref{eq:magic-substitution}) turns out to be extremely useful in
evaluating the integrals appearing here.

\subsection{Expected lifetime}
Now, that we have the asymptotic expansion of $a_0$
(eq.(\ref{eq:a_0})), the evaluation of expected lifetime (MFPT to
the absorbing boundary $\p\Omega_a$) becomes possible. Setting
$r=0$ in equations (\ref{eq:u-v}) and (\ref{eq:u-separation}), we
obtain the expression (\ref{main:mfpt-center}) for MFPT from the
center of the disk.

Averaging (\ref{eq:f-dirichlet}) with respect to a uniform initial
distribution in $\Omega$ gives
\begin{eqnarray}
E\tau & = & \frac{1}{\pi} \int_0^{2\pi}\,d\theta \int_0^1
v(r,\theta)r\,dr = \int_0^1 \left[\left(a_0 + \frac{1}{2}\right)r -
\frac{r^3}{2}\right]\,dr \nonumber \\
&&\nonumber\\
&=& \frac{a_0}{2} + \frac{1}{8} = -\log\frac{\varepsilon}{2} +
\frac{1}{8},
\end{eqnarray}
as asserted in eq.(\ref{main:averaged}).

The maximal value of the MFPT is attained at the point $r=1,
\theta=0$, which is antipodal to the center of the absorbing arc.
At this point $\ds{\frac{\partial u}{\partial \theta}}=0$, as can
be seen by differentiating expansion (\ref{eq:u-separation}) term
by term. Setting $r=1$ and $\theta=0$, we find that
\begin{equation}
v_{max} = u(1,0) = \frac{a_0}{2} + \sum_{n=1}^\infty
a_n.\label{infsum}
\end{equation}

The evaluation of the maximal exit time is not as straightforward
as the previous evaluated MFPTs, because one needs to calculate
the infinite sum in (\ref{infsum}). This calculation is done in
Appendix \ref{ap:maximal-exit-time}, where we find
(eq.(\ref{eq:v-max-appendix}))
$$
v_{max} = \log\frac{1}{\varepsilon} + 2\log 2 + O(\varepsilon),
$$
as asserted in equation (\ref{main:v_max}).

\subsection{Boundary layers} We see that the maximal exit time is only
$v_{max}-v_{center} = \log2-\ds{\frac{1}{4}}= .4431471806\ldots$
longer than its value at the center of the disk. In other words,
the variance along the radius $\theta=0, \, 0\leq r \leq 1$ is
very small. However, in the opposite direction $\theta=\pi, \,
0\leq r \leq 1$, we expect a much different behavior. In
particular, the MFPT is decreasing from a value of $v_{center}
\approx \log\ds{\frac{1}{\varepsilon}}$ at the center of the disk
to $v(1,\pi)=0$ at the center of $\p\Omega_a$. The calculation of
the exit time
\begin{equation}
v_{ray}(r)  \equiv v(r,\theta=\pi) = \frac{1-r^2}{4} + \frac{a_0}{2}
+ \sum_{n=1}^\infty a_n(-r)^n,
\end{equation}
is similar to that of the maximal exit time and is done in
Appendix \ref{ap:ray}. For $\varepsilon \ll 1$ and $1-r \gg
\sqrt{\varepsilon}$, we find the asymptotic form
(eq.(\ref{eq:v-ray-asym}))
\begin{equation}
\label{eq:v-ray-asym2} v_{ray}(r) = -\log\frac{\varepsilon}{2}
+2\log(1-r) +\frac{1-r^2}{4} - \log(1+r^2) +q(r) + O(\varepsilon),
\end{equation}
where $q(r)$ is a smooth function in the interval $[0,1]$
(eqs.(\ref{eq:q-define})-(\ref{eq:q-end-points})). Clearly, this
asymptotic expansion does not hold all the way through to the
absorbing arc at $r=1$, where the boundary condition requires
$v_{ray}(r=1)\,=0$. Instead, the boundary condition is almost
satisfied at \mbox{$r=1-\sqrt{\ds{\frac{\varepsilon}{2}}}$}
\begin{equation}
v_{ray}\left(1-\sqrt{\frac{\varepsilon}{2}}\,\right) =
-\log\frac{\varepsilon}{2}
+2\log\left(\sqrt{\frac{\varepsilon}{2}}\right) +O(\varepsilon) =
O(\varepsilon),
\end{equation}
In other words, the asymptotic series (\ref{eq:v-ray-asym2}) is
the outer expansion \cite{Bender}.

We proceed to construct the boundary layer for $1-r \ll
\sqrt{\varepsilon}$. Setting $\delta=1-r$, we have the identities
\begin{eqnarray}
\frac{1-r^2}{4} & = & \frac{1}{2}\delta -\frac{1}{4}\delta^2, \nonumber \\
&&\nonumber\\
1-2r\cos\varepsilon + r^2 & = &
4\sin^2\frac{\varepsilon}{2}\,(1-\delta) + \delta^2. \nonumber
\end{eqnarray}
The exact form of the MFPT along the ray, eq.(\ref{eq:v_ray2}),
gives the expansion
 \beq
v_{ray}(\delta)&=& \frac{\delta}{2}  + \frac{a_0 \delta}{4
\sin\ds{\frac{\varepsilon}{2}}}\label{vray}\\
&&\nonumber\\
&& -\frac{\delta}{\pi\sin\ds{\frac{\varepsilon}{2}}}
\int_0^{\cos(\varepsilon/2)}\frac{\arccos \sqrt
{s^2+\sin^2\ds{\frac{\varepsilon}{2}}}\
s^2\,ds}{\left(s^2+\sin^2\ds{\frac{\varepsilon}{2}}\right)^{3/2}}
+ O\left(\frac{\delta^2}{\varepsilon}\right).\nonumber
 \eeq
Evaluating the integral in eq.(\ref{vray}),
\begin{eqnarray}
&& \int_0^{\cos(\varepsilon/2)}\frac{\arccos \sqrt
{s^2+\sin^2\ds{\frac{\varepsilon}{2}}}\
s^2\,ds}{\left(s^2+\sin^2\ds{\frac{\varepsilon}{2}}\right)^{3/2}}\nonumber\\
&&\nonumber\\
&&=-\frac{\pi}{2}\left[\log\sin\ds{\frac{\varepsilon}{2}}+\cos\frac{\varepsilon}{2}
-\log\left(1+\cos\frac{\varepsilon}{2}\right)+\log2 \right] +
O(\varepsilon),
\end{eqnarray}
we obtain the boundary layer structure
\begin{eqnarray}
\label{eq:v_ray-boundary} v_{ray}(\delta) & = &
\frac{\delta}{\varepsilon}+
O\left(\delta,\,\frac{\delta^2}{\varepsilon}\right).
\end{eqnarray}
In particular, setting $\delta_0=-\varepsilon \log
\ds{\frac{\varepsilon}{2}}$ yields
\begin{equation}
v_{ray}(\delta_0) = -\log \frac{\varepsilon}{2} + O(\varepsilon
\log^2 \varepsilon),
\end{equation}
which is the value of the outer solution. We conclude that the
width of the boundary layer is $O\left(\varepsilon
\log\ds{\frac{1}{\varepsilon}}\right)$. Furthermore, the flux at
the center of the hole is given by
\begin{equation}
\label{eq:flux-center} \mbox{flux}_{\mbox{center}} =
\frac{\partial v_{ray}}{\partial r}\bigg|_{r=1} = -\frac{\partial
v_{ray}}{\partial \delta}\bigg|_{\delta=0} =
-\frac{1}{\varepsilon} + O(1).
\end{equation}

\subsection{Flux profile} Next, we calculate the profile of
the flux on the absorbing arc. Differentiating expansion
(\ref{eq:u-separation}) gives the flux as
\begin{equation}
f(\theta) = \frac{\partial v(r,\theta)}{\partial
r}\bigg|_{r=1}=\frac{\partial u(r,\theta)}{\partial
r}\bigg|_{r=1}-\frac{1}{2} = -\frac{1}{2}+\sum_{n=1}^\infty n a_n
\cos n\theta,
\end{equation}
for $\pi-\varepsilon < \theta \leq \pi$. Using equation
(\ref{eq:a_n}) for the coefficients, we have
\begin{eqnarray}
f(\theta) & = & -\frac{1}{2} +  \frac{1}{\sqrt{2}}
\int_0^{\pi-\varepsilon} h_1(t) \,dt\sum_{n=1}^\infty n[P_n(\cos
t) +
P_{n-1}(\cos t)]\cos n\theta \nonumber \\
& = & -\frac{1}{2} + \frac{1}{\sqrt{2}} \frac{d}{d\theta}
\int_0^{\pi-\varepsilon} h_1(t) \,dt\sum_{n=1}^\infty [P_n(\cos t)
+ P_{n-1}(\cos t)]\sin n\theta. \nonumber
\end{eqnarray}
Since $\theta \rangle  \pi-\varepsilon \rangle  t$, equation
(\ref{eq:heavyside}) implies
\begin{equation}
\label{eq:flux-h1} f(\theta) = -\frac{1}{2} + \frac{d}{d\theta}
\left(\cos\frac{\theta}{2}\int_0^{\pi-\varepsilon}
\frac{h_1(t)\,dt}{\sqrt{\cos t - \cos \theta}} \right).
\end{equation}
The evaluation of this integral is not immediate and is given in
Appendix \ref{ap:flux}. We find that
(eq.(\ref{eq:flux-boundary-expansion}))
\begin{eqnarray}
f(\alpha) & = & -\frac{\alpha^2}{\varepsilon \sqrt{1-\alpha^2}} -
\frac{1}{\varepsilon} \sum_{n=0}^\infty
\left(\frac{\left(2^{n+1}(n+1)!\right)^2}{(2n+2)!}\alpha^2-\frac{(2^n
n!)^2}{(2n+1)!} \right)(1 - \alpha^2)^{n+1/2} \nonumber \\
&&\nonumber\\ & & - \frac{\pi}{2\varepsilon}\sum_{n=0}^\infty
\left(\frac{(2n)!}{(2^n n!)^2
}-\frac{(2n+2)!(2n+2)}{(2^{n+1}(n+1)!)^2}\alpha^2
\right)(1-\alpha^2)^n + O(1),
\end{eqnarray}
where $\alpha = \ds{\frac{\pi-\theta}{\varepsilon}},\,
|\alpha|<1$. The flux has a singular part, represented by the
half-integer powers of $(1-\alpha^2$), and a remaining regular
part (the integer powers.) The first term,
$-\ds{\frac{\alpha^2}{\varepsilon \sqrt{1-\alpha^2}}}$, is the
most singular one, because it becomes infinite as
$|\alpha|\rightarrow 1$. In other words, the flux is infinitely
large near the boundary of the hole. The splitting of the solution
into singular and regular parts is common in the theory of
elliptic boundary value problems in domains with corners (see
e.g., \cite{Mazya1}-\cite{Dauge}).

The value of the flux at the center of the hole is to leading order
\begin{eqnarray}
\label{eq:flux-center2} f(0) & = &  -\frac{1}{\varepsilon}
\sum_{n=0}^\infty \left(\frac{\pi}{2}\frac{(2n)!}{(2^n n!)^2 } -
\frac{(2^n n!)^2}{(2n+1)!}\right) = -\frac{1}{\varepsilon},
\end{eqnarray}
in agreement with (\ref{eq:flux-center}) (thanks Maple for
calculating the infinite sum.)

The size of the boundary layer is varying with $\theta$
proportionally to $1/f(\theta)$. The singularity at the end points
of the hole indicate that the layer shrinks there to zero.
Therefore, the boundary layer is shaped as a small cap bounded by
the absorbing arc and (more or less) the curve $\sqrt{1-\alpha^2}$
(see Fig.\ref{f:boundary-layer}). In particular, the MFPT on the
reflecting boundary is
$O\left(\log\ds{\frac{1}{\varepsilon}}\right)$, even when taken
arbitrarily close to the absorbing boundary. The singularity of
the flux near the endpoints indicates that the diffusive particle
prefers to exit near the endpoints rather than through the center
of the hole.

The expansion (\ref{eq:flux-boundary-expansion}) is useful in
approximating the flux near the endpoints $(\alpha=\pm 1)$, where
few terms are needed. However, it is slowly converging near the
center of the hole, where a power series in $\alpha^2$ should be
used instead
\begin{equation}
f(\alpha) = \sum_{n=0}^\infty f_{n} \alpha^{2n} + O(1),
\end{equation}
where the coefficients $f_n$ are $O(\varepsilon^{-1})$. Equations
(\ref{eq:flux-center2}) and (\ref{eq:flux-center}) indicate that
$f_0=-\ds{\frac{1}{\varepsilon}}$. All other coefficients can be
found in a similar fashion. We conclude that near the center
$(\alpha \ll 1)$ we have
\begin{equation}
f(\alpha) = -\frac{1}{\varepsilon} + O\left(1,
\frac{\alpha^2}{\varepsilon}\right).
\end{equation}

\appendix
\section{Maximal exit time for the circular disk}
\label{ap:maximal-exit-time}
Using equation (\ref{eq:a_n}) we find
\begin{eqnarray}
v_{max} & = & u(1,0) = \frac{a_0}{2} + \sum_{n=1}^\infty a_n
\\
& = & \frac{a_0}{2} +  \frac{1}{\sqrt{2}} \int_0^{\pi-\varepsilon}
h_1(t) \sum_{n=1}^\infty[P_n(\cos t) + P_{n-1}(\cos t)]\,dt.
\nonumber
\end{eqnarray}
Recall the generating function of the Legendre polynomials
\cite{Stegun}
\begin{equation}
\label{eq:generating} \frac{1}{\sqrt{1-2tx+t^2}} = \sum_{n=0}^\infty
P_n(x)t^n,
\end{equation}
from which it follows that
\begin{equation}
\sum_{n=0}^\infty P_n(\cos t) = \frac{1}{\sqrt{1-2\cos t+1}} =
\frac{1}{2\sin\ds{\frac{t}{2}}}.
\end{equation}
Together with equation (\ref{a_0}), this gives
\begin{equation}
v_{max} = \frac{a_0}{2} + \frac{1}{\sqrt{2}}
\int_0^{\pi-\varepsilon} h_1(t)
\left(\frac{1}{\sin\ds{\frac{t}{2}}}-1\right)\,dt =
\frac{1}{\sqrt{2}} \int_0^{\pi-\varepsilon}
\frac{h_1(t)\,dt}{\sin\ds{\frac{t}{2}}}.
\end{equation}
Combining with equation (\ref{eq:h_1}) and integrating by parts,
we get
\begin{eqnarray}
v_{max}&=&\frac{1}{\sqrt{2}} \int_0^{\pi-\varepsilon}
\frac{1}{\pi} \frac{1}{\sin\ds{\frac{t}{2}}} \frac{d}{dt} \int_0^t
\frac{u \sin\ds{
\frac{u}{2}}\,du}{\sqrt{\cos u - \cos t}}\,dt \\
&&\nonumber\\
&=& \frac{1}{\sqrt{2}\pi \sin\ds{\frac{t}{2}}} \int_0^t
\frac{u\sin\ds{\frac{u}{2}}\,du}{\sqrt{\cos u - \cos
t}}\bigg|_0^{\pi-\varepsilon}  \nonumber\\
&&\nonumber\\
&&+\frac{1}{2\sqrt{2}\pi} \int_0^{\pi-\varepsilon}
\frac{\cos\ds{\frac{t}{2}}}{\sin^2\ds{\frac{t}{2}}}\,dt \int_0^t
\frac{u\sin\ds{\frac{u}{2}}\,du}{\sqrt{\cos u - \cos t}}
.\nonumber
\end{eqnarray}
Equations (\ref{eq:int}) and (\ref{eq:a_0}) show that
\begin{equation}
\frac{\sqrt{2}}{\pi} \int_0^{t} \frac{u \sin\ds{
\frac{u}{2}}\,du}{\sqrt{\cos u - \cos t}} = -2\log \cos
\frac{t}{2}+ 2\log\left(1+\sin\ds{\frac{t}{2}} \right) + k(t),
\end{equation}
where
\begin{equation}
k(t) = - \frac{4}{\pi} \int_0^{\ds{\sin\frac{t}2}}
\left(\frac{\arcsin \sqrt{s^2+\cos^2
{\frac{t}{2}}}}{\sqrt{s^2+\cos^2\ds{ \frac{t}{2}}}}\right)\,ds.
\end{equation}
Therefore,
\begin{eqnarray}
\lim_{t\rightarrow 0} \frac{\sqrt{2}}{\pi
\sin\ds{\frac{t}{2}}}\int_0^{t} \frac{u \sin\ds{
\frac{u}{2}}\,du}{\sqrt{\cos u - \cos t}} & = & \lim_{t\rightarrow
0} \frac{-2\log \cos\ds{\frac{t}{2}}+
2\log\left(1+\sin\ds{\frac{t}{2}}
\right) + k(t)}{\sin\ds{\frac{t}{2}}} \nonumber \\
&&\nonumber\\
& = & 2 - \frac{4}{\pi}\arcsin(1) = 0.
\end{eqnarray}
Hence
\begin{eqnarray*}
&&v_{max} =\\
&&\\
&& \frac{1}{2\cos\ds{\frac{\varepsilon}{2}}} \left[
2\log\left(1+\ds{\cos\frac{\varepsilon}{2}}\right) -
2\log\sin\ds{\frac{\varepsilon}{2}} -
\frac{4}{\pi}\int_0^{\ds{\cos\frac{\varepsilon}{2}}}\left(\frac{\arcsin
\sqrt{s^2+\sin^2\ds{\frac{\varepsilon}{2}}}}{\sqrt{s^2+\sin^2\ds{
\frac{\varepsilon}{2}}}}\right)\,ds \right] \nonumber \\
&&\\
 && + \frac{1}{2\sqrt{2}\pi} \int_0^{\pi-\varepsilon}
\frac{\cos\ds{\frac{t}{2}}}{\sin^2\ds{\frac{t}{2}}}\,dt \int_0^t
\frac{u\sin\ds{\frac{u}{2}}\,du}{\sqrt{\cos u - \cos t}}.
\end{eqnarray*}
For $\varepsilon \ll 1$
\begin{equation}
v_{max} = -\log\frac{\varepsilon}{2} + \frac{1}{2\sqrt{2}\pi}
\int_0^{\pi}
\frac{\cos\ds{\frac{t}{2}}}{\sin^2\ds{\frac{t}{2}}}\,dt \int_0^t
\frac{u\sin\ds{\frac{u}{2}}\,du}{\sqrt{\cos u - \cos t}} +
O(\varepsilon).
\end{equation}
Changing the order of integration, we get
\begin{equation}
v_{max} = -\log\frac{\varepsilon}{2} + \frac{1}{2\sqrt{2}\pi}
\int_0^{\pi} u\sin\ds{\frac{u}{2}}\,du \int_u^\pi
\frac{\cos\ds{\frac{t}{2}}\,dt}{\sin^2\ds{\frac{t}{2}}\sqrt{\cos u
- \cos t}} + O(\varepsilon).
\end{equation}
Substituting
\begin{equation}
\label{eq:magic-substitution2} s=\sqrt{\frac{\cos u - \cos t}{2}}
\end{equation}
in the inner integral results in
\begin{eqnarray*}
\int_u^\pi
\frac{\cos\ds{\frac{t}{2}}\,dt}{\sin^2\ds{\frac{t}{2}}\sqrt{\cos u
- \cos t}} = \sqrt{2}
\frac{\cos\ds{\frac{u}{2}}}{\sin^2\ds{\frac{u}{2}}}. \nonumber
\end{eqnarray*}
Therefore,
\begin{eqnarray}
v_{max}
 & = & -\log\frac{\varepsilon}{2} + \frac{1}{2\pi} \int_0^\pi
\frac{u}{\tan\ds{\frac{u}{2}}}\,du = -\log\frac{\varepsilon}{2} -
\frac{2}{\pi} \int_0^{\pi/2} \log \sin v \, dv \nonumber \\
&&\nonumber\\
 & = & -\log\frac{\varepsilon}{2} + \log2.
\label{eq:v-max-appendix}
\end{eqnarray}

\section{Exit times along the ray}
\label{ap:ray} Along the ray $\theta=\pi$ the MFPT is given by
\begin{eqnarray*}
v_{ray}(r) & \equiv & v(r,\theta=\pi) = \frac{1-r^2}{4} +
\frac{a_0}{2} +
\sum_{n=1}^\infty a_n(-r)^n \\
&&\\ & = & \frac{1-r^2}{4} + \frac{a_0}{2} + \frac{1}{\sqrt{2}}
\int_0^{\pi-\varepsilon} h_1(t) \sum_{n=1}^\infty [P_n(\cos t) +
P_{n-1}(\cos t)](-r)^n\,dt.
 \nonumber
\end{eqnarray*}
Using the generating function (\ref{eq:generating}) of the
Legendre polynomials to sum the infinite series, we obtain
\begin{equation}
v_{ray}(r) = \frac{1-r^2}{4} + \frac{1-r}{\sqrt{2}}
\int_0^{\pi-\varepsilon} \frac{h_1(t)\,dt}{\sqrt{1+2r\cos t +
r^2}}.
\end{equation}
Combining with equation (\ref{eq:h_1}), integrating by parts, and
hanging the order of integration gives
\begin{eqnarray*}
&&v_{ray}(r) = \frac{1-r^2}{4} + \frac{1-r}{2 \sqrt{1-2r\cos
\varepsilon +r^2}}\, a_0 \\
&&\\
& & - \frac{r(1-r)}{\sqrt{2}\pi}  \int_0^{\pi-\varepsilon}
u\sin\ds{\frac{u}{2}}\,du \int_u^{\pi-\varepsilon} \frac{\sin t
\,dt}{(1+2r\cos t + r^2)^{3/2}\sqrt{\cos u - \cos t}}. \nonumber
\end{eqnarray*}
The substitutions $s=\sqrt{\cos u - \cos t}$ and $x=\sqrt{2r}\,s$
lead to
\begin{eqnarray*}
\int_u^{\pi-\varepsilon} \frac{\sin t \,dt}{(1+2r\cos t +
r^2)^{3/2}\sqrt{\cos u - \cos t}} = \frac{2\sqrt{\cos u + \cos
\varepsilon}}{(1+2r\cos u + r^2)\sqrt{1-2r\cos \varepsilon +
r^2}},
\end{eqnarray*}
which implies that
\begin{eqnarray}
\label{eq:v_ray} v_{ray}(r) & = & \frac{1-r^2}{4} + \frac{1-r}{2
\sqrt{1-2r\cos \varepsilon +
r^2}}\, a_0 \\
&&\nonumber\\
 & & -
\frac{\sqrt{2}\,r(1-r)}{\pi\sqrt{1-2r\cos\varepsilon+r^2}}
\int_0^{\pi-\varepsilon} u\sin\ds{\frac{u}{2}}\frac{\sqrt{\cos u +
\cos \varepsilon}}{1+2r\cos u + r^2}\,du .\nonumber
\end{eqnarray}
The substitution (\ref{eq:magic-substitution}) gives
\begin{eqnarray*}
&&\int_0^{\pi-\varepsilon} u\sin\ds{\frac{u}{2}}\frac{\sqrt{\cos u
+ \cos \varepsilon}}{1+2r\cos u + r^2}\,du =\\
&&\\
&& 4\sqrt{2} \int_0^{\ds{\cos\frac\varepsilon2}} \frac{\arccos
\sqrt{s^2+\sin^2\ds{\frac{\varepsilon}{2}}}\, s^2 \,
ds}{(1-2r\cos\varepsilon+r^2+4rs^2)\sqrt{\sin^2\ds{\frac{\varepsilon}{2}}+s^2}\,},
\end{eqnarray*}
and we obtain the exact form of $v_{ray}(r)$ as
\begin{eqnarray}
&& v_{ray}(r) = \frac{1-r^2}{4} + \frac{1-r}{2 \sqrt{1-2r\cos
\varepsilon +
r^2}}\, a_0\label{eq:v_ray2} \\
&&\nonumber\\
& & - \frac{8r(1-r)}{\pi\sqrt{1-2r\cos\varepsilon+r^2}}
\int_0^{\ds{\cos\frac\varepsilon2}} \frac{\arccos
\sqrt{s^2+\sin^2\ds{\frac{\varepsilon}{2}}}\, s^2 \,
ds}{(1-2r\cos\varepsilon+r^2+4rs^2)\sqrt{\sin^2\ds{\frac{\varepsilon}{2}}+s^2}}.
\nonumber
\end{eqnarray}
For $\varepsilon \ll 1$ and $1-r \gg \sqrt{\varepsilon}$ equation
(\ref{eq:v_ray2}) becomes
\begin{equation}
v_{ray}(r) = \frac{1-r^2}{4} -\log\frac{\varepsilon}{2} -
\frac{8r}{\pi} \int_0^1 \frac{s\arccos s  \, ds}{(1-r)^2+4rs^2} +
O(\varepsilon).\label{vrayr}
\end{equation}
To evaluate the integral in (\ref{vrayr}), we write
\begin{equation}
\label{eq:arccos} \arccos s = \frac{\pi}{2} - \arcsin s,
\end{equation}
and obtain
\begin{eqnarray*}
\frac{8r}{\pi}\frac{\pi}{2} \int_0^{1} \frac{s \,
ds}{(1-r)^2+4rs^2}  =  -2\log(1-r) + \log(1+r^2).
\end{eqnarray*}
The function $q(r)$, defined by
\begin{equation}
\label{eq:q-define} q(r)=\frac{8r}{\pi} \int_0^{1} \frac{\arcsin s
\,s \, ds}{(1-r)^2+4rs^2}
\end{equation}
in the interval $0\leq r\leq1$, has the endpoint values
\begin{equation}
\label{eq:q-end-points} q(0) = 0, \quad q(1) = \log2.
\end{equation}
Therefore,
\begin{equation}
\label{eq:v-ray-asym} v_{ray}(r) = -\log\frac{\varepsilon}{2}
+2\log(1-r) +\frac{1-r^2}{4} - \log(1+r^2) + q(r) +
O(\varepsilon),
\end{equation}
is the MFPT for $\varepsilon \ll 1$ and $1-r \gg
\sqrt{\varepsilon}$. In particular,
 $$v_{center}=v_{ray}(0) =
-\log\frac{\varepsilon}{2}+\frac{1}{4}+O(\varepsilon),$$ as
asserted in (\ref{main:mfpt-center}).

\section{Flux profile}
\label{ap:flux} In this appendix we calculate the flux profile given
by equation (\ref{eq:flux-h1}). Substituting equation (\ref{eq:h_1})
for $h_1$ in equation (\ref{eq:flux-h1}) gives
 \[
f(\theta)  =  -\ds{\frac{1}{2}} + \ds{\frac{d}{d\theta}}
\left[\cos\frac{\theta}{2}\ds{\int_0^{\pi-\varepsilon}}
\ds{\frac{\ds{\frac{1}{\pi} \frac{d}{dt}} \ds{\int_0^t \frac{u
\sin\ds{\frac{u}{2}}\,du}{\sqrt{\cos u - \cos t}}}}{\sqrt{\cos t -
\cos \theta}}} \,dt\right].
 \]
Integration by parts and changing the order of integration, we
find that
\begin{eqnarray*}
f(\theta)  & = & -\frac{1}{2} + \frac{1}{\pi}\frac{d}{d\theta}
\Biggl[ \frac{\cos\ds{\frac{\theta}{2}}}{\sqrt{\cos
(\pi-\varepsilon) - \cos \theta}}\int_0^{\pi-\varepsilon}
\frac{u\sin\ds{\frac{u}{2}}\,du}{\sqrt{\cos u + \cos \varepsilon}} \\
&&\\
 & & -\frac{1}{2} \cos\ds{\frac{\theta}{2}}
\int_0^{\pi-\varepsilon} u \sin\ds{\frac{u}{2}}\,du
\int_u^{\pi-\varepsilon} \frac{\sin t \,dt}{(\cos t - \cos
\theta)^{3/2}(\cos u - \cos t)^{1/2}}\Biggl]. \nonumber
\end{eqnarray*}
We evaluate the inner integral by making the substitution
$x=\sqrt{\cos u - \cos t}$,
\begin{eqnarray*}
&&\int_u^{\pi-\varepsilon} \frac{\sin t \,dt}{(\cos t - \cos
\theta)^{3/2}(\cos u - \cos t)^{1/2}} = \frac{2\sqrt{\cos u + \cos
\varepsilon}}{(\cos u - \cos \theta)\sqrt{-\cos \theta - \cos
\varepsilon}}.
\end{eqnarray*}
Therefore
 \beqq
f(\theta) & = & -\frac{1}{2} + \frac{1}{\pi}\frac{d}{d\theta}
\Biggl[ \frac{\cos\ds{\frac{\theta}{2}}}{\sqrt{-\cos \varepsilon -
\cos \theta}}\int_0^{\pi-\varepsilon}
\frac{u\sin\ds{\frac{u}{2}}\,du}{\sqrt{\cos u + \cos \varepsilon}} \nonumber \\
&&\\
 & & - \frac{\cos\ds{\frac{\theta}{2}}}{\sqrt{-\cos \theta - \cos
\varepsilon}} \int_0^{\pi-\varepsilon} \frac{u
\sin\ds{\frac{u}{2}}\sqrt{\cos u + \cos \varepsilon}}{\cos u -
\cos
\theta}\,du \Biggl] \\
&&\\
 & = & -\frac{1}{2} + \frac{1}{\pi}\frac{d}{d\theta} \left[
\frac{\cos\ds{\frac{\theta}{2}}}{\sqrt{-\cos \varepsilon - \cos
\theta}}\frac{\pi}{\sqrt{2}}a_0
-\right.\nonumber\\
&&\nonumber\\
&&\left.\frac{\cos\ds{\frac{\theta}{2}}}{\sqrt{-\cos \theta - \cos
\varepsilon}} \int_0^{\pi-\varepsilon} \frac{u
\sin\ds{\frac{u}{2}}\sqrt{\cos u + \cos \varepsilon}}{\cos u -
\cos \theta}\,du \right].\nonumber
 \eeqq
The substitution (\ref{eq:magic-substitution}) gives
 \beqq
&&\int_0^{\pi-\varepsilon} \frac{u \sin\ds{\frac{u}{2}} \sqrt{\cos
u + \cos \varepsilon}\,du}{\cos u - \cos \theta} =\\
&&\\
&& 2\sqrt{2} \int_0^{\ds{\cos\frac\varepsilon2}}\frac{\arccos
\sqrt{s^2+\sin^2\ds{\frac{\varepsilon}{2}}}\,s^2\,ds}
{\left(s^2+\sin^2\ds{\frac{\varepsilon}{2}}-\cos^2
\frac{\theta}{2}\right)\sqrt{s^2+\sin^2\ds{\frac{\varepsilon}{2}}}\,}.
 \eeqq
Therefore, the flux takes the form
 \beqq
&&f(\theta)  = -\frac{1}{2}
+\frac{1}{\sqrt{2}\pi}\frac{d}{d\theta} \left[
\frac{\cos\ds{\frac{\theta}{2}}}{\sqrt{-\cos \theta - \cos
\varepsilon}}\times\right.\\
&&\\
&&\left. \left(\pi a_0 - \int_0^{\cos(\varepsilon/2)}
\frac{4\arccos \sqrt{s^2+\sin^2\ds{\frac{\varepsilon}{2}}}\,
s^2\,ds}{\left(s^2+\sin^2\ds{\frac{\varepsilon}{2}}-\cos^2\ds{\frac{\theta}{2}}\right)
\sqrt{\sin^2\ds{\frac{\varepsilon}{2}}+s^2}}\right) \right],
 \eeqq
which is rewritten as
\begin{equation}
\label{eq:flux-integral-phi} f(\theta) =  -\frac{1}{2} +
\frac{1}{2\pi}\frac{d}{d\theta} \left[
\frac{\cos\ds{\frac{\theta}{2}}}{b} \left(\pi a_0 - 4
\int_0^{\sqrt{1-a^2}} \frac{\arccos \sqrt{s^2+a^2}\,
s^2\,ds}{\sqrt{a^2+s^2} \,\left(s^2+b^2 \right)}\right) \right],
\end{equation}
where $a=\sin\ds{\frac{\varepsilon}{2}}$ and
$2b^2=-\cos\theta-\cos\varepsilon$. Writing
 \beqq
 \phi(a,s) =\ds{\frac{\arccos\sqrt{s^2+a^2}}{\sqrt{s^2+a^2}}} = \sum_{n=0}^\infty \phi_{2n}(a)\, s^{2n},
 \eeqq
we find the Taylor coefficients
 $$\phi_0(a) = \ds{\frac{\arccos
a}{a}},\;\phi_2(a) = -\left(\ds{\frac{\arccos
 a}{2a^3}}+\ds{\frac{1}{2a^2\sqrt{1-a^2}}} \right),$$
and so on. For all $n \geq 0$ we find the asymptotic behavior
 $$\phi_{2n}(a) \sim \ds{\frac{c_{2n}}{a^{2n+1}}} +
 O\left(\ds{\frac{1}{a^{2n}}}\right)\quad\mbox{as}\quad a \rightarrow 0.$$
To see this, consider the Taylor expansions
\begin{eqnarray*}
\left(\sqrt{1+\left(\frac{s}{a}\right)^2}\,\right)^{2n+1} & = &
\sum_{m=0}^\infty
c_{n}^m \frac{s^{2m}}{a^{2m}} \nonumber \\
&&\\
 \arccos\left(a\sqrt{1+\left(\frac{s}{a}\right)^2}\right) & = &
\frac{\pi}{2} + \sum_{n=0}^\infty \alpha_n a^{2n+1}
\left(\sqrt{1+\left(\frac{s}{a} \right)^2}\,\right)^{2n+1} \nonumber \\
&&\\
 & = & \frac{\pi}{2} + \sum_{n=0}^\infty \alpha_n a^{2n+1}
\sum_{m=0}^\infty
c_{n}^m \frac{s^{2m}}{a^{2m}} \nonumber \\
&&\\ & = & \frac{\pi}{2} + \sum_{m=0}^\infty
\frac{s^{2m}}{a^{2m}}\sum_{n=0}^\infty c_n^m \alpha_n a^{2n+1},
 \eeqq
where $\alpha_n$ and $c_n^m$ are (known) constants, and
\begin{equation}
\frac{1}{a\sqrt{1+\left(\ds{\frac{s}{a}}\right)^2}} =
\ds{\frac{1}{a}} \sum_{n=0}^\infty (-1)^n \frac{(2n)!}{(2^n
n!)^2}\frac{s^{2n}}{a^{2n}}.
\end{equation}
Therefore
\begin{equation}
\phi(a,s) = \frac{1}{a} \sum_{n=0}^\infty (-1)^n \frac{(2n)!}{(2^n
n!)^2}\frac{s^{2n}}{a^{2n}} \left(\frac{\pi}{2} + \sum_{m=0}^\infty
\frac{s^{2m}}{a^{2m}}\sum_{n=0}^\infty c_n^m \alpha_n a^{2n+1}
\right),
\end{equation}
from which it follows that
\begin{equation}
\label{eq:phi-coeff-asym} \phi(a,s) = \sum_{n=0}^\infty
\left((-1)^n \frac{\pi}{2}\frac{(2n)!}{(2^n n!)^2}+O(a)
\right)\frac{s^{2n}}{a^{2n+1}}.
\end{equation}
This shows that
 \beq
 \phi_{2n}(a) \sim (-1)^n
\frac{\pi}{2}\frac{(2n)!}{(2^n n!)^2a^{2n+1}} + O\left(a^{-2n}
\right)\label{phi2n},
 \eeq
as asserted. The asymptotic behavior (\ref{phi2n}) of the
coefficients $\phi_{2n}(a)$ can be used to estimate the integral
in equation (\ref{eq:flux-integral-phi}),
\begin{equation}
\label{eq:int-for-phi} \int_0^{\sqrt{1-a^2}}
\frac{\phi(a,s)\,s^2\,ds}{s^2+b^2} = \sum_{n=0}^\infty
\phi_{2n}(a)\int_0^{\sqrt{1-a^2}} \frac{s^{2n+2}\,ds}{s^2+b^2}.
\end{equation}
To extract the asymptotic behavior of the integral as $b\to0$, we
use the long division
\begin{equation}
\frac{s^{2n+2}}{s^2+b^2} = \sum_{j=0}^n (-1)^j b^{2j} s^{2n-2j} +
\frac{(-1)^{n+1} b^{2n+2}}{s^2+b^2}
\end{equation}
and integrate it to yield
\begin{eqnarray}
&&\int_0^{\sqrt{1-a^2}}\frac{s^{2n+2}}{s^2+b^2}\,ds =\label{eq:yet-another-intergal}\\
&&\nonumber\\
&& \sum_{j=0}^n\left[ (-1)^j b^{2j}
\int_0^{\sqrt{1-a^2}}s^{2n-2j}\,ds \right] + (-1)^{n+1}
b^{2n+2}\int_0^{\sqrt{1-a^2}}\frac{ds}{s^2+b^2}
\nonumber \\
&&\nonumber\\
&  &= \sum_{j=0}^n \left[ (-1)^j b^{2j}
\frac{(\sqrt{1-a^2})^{2n-2j+1}}{2n-2j+1}\right] + (-1)^{n+1}
b^{2n+1}\arctan\frac{\sqrt{1-a^2}}{b}. \nonumber
\end{eqnarray}
The Taylor expansion
\begin{equation}
\arctan\frac{\sqrt{1-a^2}}{b} = \frac{\pi}{2} + \sum_{m=0}^\infty
\frac{(-1)^{m+1}}{2m+1}\frac{b^{2m+1}}{(\sqrt{1-a^2})^{2m+1}}
\end{equation}
gives the Taylor expansion of the integral
(\ref{eq:yet-another-intergal}) in powers of $b$ as
\begin{eqnarray*}
\int_0^{\sqrt{1-a^2}}\frac{s^{2n+2}}{s^2+b^2}\,ds & = &
\sum_{j=0}^n (-1)^j b^{2j}
\frac{(\sqrt{1-a^2})^{2n-2j+1}}{2n-2j+1}
\nonumber \\
&&\\
 & & + (-1)^{n+1} b^{2n+1}\left(\frac{\pi}{2} +
\sum_{m=0}^\infty
\frac{(-1)^{m+1}}{2m+1}\frac{b^{2m+1}}{(\sqrt{1-a^2})^{2m+1}}
\right) \nonumber \\
&&\\
 & = & \sum_{j=0}^n (-1)^j
\frac{(\sqrt{1-a^2})^{2n-2j+1}}{2n-2j+1}\,b^{2j}\\
&&\\
& & + (-1)^{n+1}\frac{\pi}{2}\, b^{2n+1} +  \sum_{m=0}^\infty
\frac{(-1)^{n+m}}{(2m+1)(\sqrt{1-a^2})^{2m+1}}\,b^{2m+2n+2}.
 \nonumber
\end{eqnarray*}
Therefore, the Taylor expansion of the integral
(\ref{eq:int-for-phi}) is
\begin{eqnarray*}
&&\int_0^{\sqrt{1-a^2}} \frac{\phi(a,s)\,s^2\,ds}{s^2+b^2}=
\sum_{n=0}^\infty \phi_{2n}(a) \biggl[\sum_{j=0}^n (-1)^j
\frac{(\sqrt{1-a^2})^{2n-2j+1}}{2n-2j+1}\,b^{2j}\\
&&\\
& & + (-1)^{n+1}\frac{\pi}{2}\, b^{2n+1} +  \sum_{m=0}^\infty
\frac{(-1)^{n+m}}{(2m+1)(\sqrt{1-a^2})^{2m+1}}\,b^{2m+2n+2}
\biggr]. \nonumber
\end{eqnarray*}
Rearranging in powers of $b$, we find that
\begin{equation}
\int_0^{\sqrt{1-a^2}} \frac{\phi(a,s)\,s^2\,ds}{s^2+b^2} =
\sum_{n=0}^\infty \beta_n(a)\, b^n,
\end{equation}
where the first three coefficients are
\begin{eqnarray*}
\beta_0(a) & = & \sum_{n=0}^\infty \phi_{2n}(a)
\frac{(\sqrt{1-a^2})^{2n+1}}{2n+1}
 = \int_0^{\sqrt{1-a^2}}\phi(a,s)\,ds = \frac{\pi}{4}\,a_0, \nonumber \\
 &&\\
 \beta_1(a) & = & -\frac{\pi\phi_0(a)}{2} = -\frac{\pi \arccos a}{2a}, \nonumber \\
 &&\\
\beta_2(a) & = & -\sum_{n=1}^\infty \phi_{2n}(a)
\frac{(\sqrt{1-a^2})^{2n-1}}{2n-1} +
\frac{\phi_0(a)}{\sqrt{1-a^2}}
\nonumber \\
&&\\
& = & -\int_0^{\sqrt{1-a^2}}\frac{\phi(a,s)-\phi_0(a)}{s^2}\,ds +
\frac{\phi_0(a)}{\sqrt{1-a^2}}
\end{eqnarray*}
and all other coefficients $\beta_n$ are recovered in a similar
fashion,
\begin{eqnarray*}
\beta_{2j+1} & = & (-1)^{j+1} \frac{\pi}{2}\phi_{2j}(a) = -
\frac{\pi^2}{4} \frac{(2j)!}{(2^j
j!)^2a^{2j+1}} + O(a^{-2j}), \nonumber \\
&&\\
 \beta_{2j} & = & (-1)^j \left(\sum_{n=j}^\infty \phi_{2n}(a)
\frac{(\sqrt{1-a^2})^{2n-2j+1}}{2n-2j+1}\right.\\
&&\\
&&\left. - \sum_{n=0}^{j-1} \phi_{2n}(a)
\frac{1}{(2j-2n-1)(\sqrt{1-a^2})^{2j-2n-1}}\right)
\nonumber \\
&&\\
& = & (-1)^j
\left(\int_0^{\sqrt{1-a^2}}\frac{1}{s^{2j}}\sum_{n=j}^\infty
\phi_{2n}(a) s^{2n}\,ds\right.\\
&&\\
&&\left. - \sum_{n=0}^{j-1} \phi_{2n}(a)
\frac{1}{(2j-2n-1)(\sqrt{1-a^2})^{2j-2n-1}}\right)
\nonumber \\
&&\\
& = & (-1)^j \left( \int_0^{\sqrt{1-a^2}}
\frac{\phi(a,s)-\sum_{n=0}^{j-1}\phi_{2n}(a)s^{2n}}{s^{2j}}\,ds\right.\\
&&\\
&&\left. - \sum_{n=0}^{j-1} \phi_{2n}(a)
\frac{1}{(2j-2n-1)(\sqrt{1-a^2})^{2j-2n-1}}\right).
\end{eqnarray*}
We see that extra effort should be put in finding the even
coefficients $\beta_{2n}$. Expanding
\begin{equation}
\phi(a,s) = \frac{\pi}{2}\frac{1}{\sqrt{s^2+a^2}} -
\sum_{n=0}^\infty \frac{(2n)!}{(2^n n!)^2}\frac{(s^2+a^2)^n}{2n+1},
\end{equation}
and noting that the following infinite sum has a regular
contribution
\begin{equation}
\lim_{a\rightarrow 0}\int_0^{\sqrt{1-a^2}}
\frac{1}{s^{2j}}\sum_{n=j}^\infty \frac{(2n)!}{(2^n
n!)^2}\frac{(s^2+a^2)^n}{2n+1} \,ds = C_j,
\end{equation}
where $C_j$ are constants (also can be written in term of
hypergeometric functions), we find an alternative representation
for the even coefficients,
\begin{eqnarray*}
\beta_{2j} & = & (-1)^j \left( \int_0^{\sqrt{1-a^2}}
\frac{\phi(a,s)-\sum_{n=0}^{j-1}\phi_{2n}(a)s^{2n}}{s^{2j}}\,ds\right.\\
&&\\
&&\left. - \sum_{n=0}^{j-1}
\frac{\phi_{2n}(a)}{(2j-2n-1)(\sqrt{1-a^2})^{2j-2n-1}}\right)\nonumber \\
&&\\
 & = & (-1)^j\biggl(-C_j + O(a) + \\
 &&\\
 &&\int_0^{\sqrt{1-a^2}}
\frac{\ds{\frac{\pi}{2}}\ds{\frac{1}{\sqrt{s^2+a^2}}}- \sum_{n=0}^{j-1}\ds{\frac{(2n)!}{(2^nn!)^2}}
\ds{\frac{(s^2+a^2)^n}{2n+1}}-\sum_{n=0}^{j-1}\phi_{2n}(a)s^{2n}}{s^{2j}}\,ds\nonumber \\
&&\\
 & & - \sum_{n=0}^{j-1} \phi_{2n}(a)
\frac{1}{(2j-2n-1)(\sqrt{1-a^2})^{2j-2n-1}}\biggr)
\nonumber \\
&&\\
 & = & (-1)^j\biggl(\int_0^{\sqrt{1-a^2}}
\frac{\ds{\frac{\pi}{2}}\ds{\frac{1}{\sqrt{s^2+a^2}}} -
\sum_{n=0}^{j-1} \ds{\frac{(2n)!}{(2^n
n!)^2}}\ds{\frac{(s^2+a^2)^n}{2n+1}}-\sum_{n=0}^{j-1}\phi_{2n}(a)s^{2n}}{s^{2j}}\,ds
\nonumber \\
&&\\
& & -
(-1)^{j-1}\frac{\pi}{2}\frac{(2j-2)!}{(2^{j-1}(j-1)!)^2}\frac{1}{a^{2j-1}}
+
O\left(\frac{1}{a^{2j-2}}\right)\biggr).\label{eq:even-beta-coeffs-integral}
\end{eqnarray*}
The integrals are given in \cite{integral},
\begin{equation}
\int \frac{ds}{s^{2j}\sqrt{s^2+a^2}} =
\frac{1}{a^{2j}}\sum_{n=0}^{j-1}\frac{(-1)^n}{2n-2j+1} {j-1 \choose
n} \left(\frac{s^2}{s^2+a^2} \right)^{n-j+1/2}.
\end{equation}
The binomial expansion gives
\begin{equation}
\int \frac{(s^2+a^2)^n\,ds}{s^{2j}} = \sum_{k=0}^n
\frac{1}{2k-2j+1}{n \choose k} s^{2k-2j+1} a^{2n-2k}.
\end{equation}
Altogether, we find that the integral term in equation
(\ref{eq:even-beta-coeffs-integral}) is
\begin{eqnarray*}
&&
\int_0^{\sqrt{1-a^2}}\frac{\ds{\frac{\pi}{2}}\ds{\frac{1}{\sqrt{s^2+a^2}}}
- \sum_{n=0}^{j-1} \ds{\frac{(2n)!}{(2^n
n!)^2}}\ds{\frac{(s^2+a^2)^n}{2n+1}}-\sum_{n=0}^{j-1}\phi_{2n}(a)s^{2n}}{s^{2j}}\,ds
= \nonumber \\
&&\\
& = &
\frac{\pi}{2}\frac{1}{a^{2j}}\sum_{n=0}^{j-1}\frac{(-1)^n}{2n-2j+1}
{j-1 \choose n} + O\left(\frac{1}{a^{2j-1}}\right).
\end{eqnarray*}
This sum has the closed form \cite{integral}
\begin{equation}
\sum_{i=0}^k \frac{(-1)^i}{2i+1}{k \choose i} = \frac{(2^k
k!)^2}{(2k+1)!},
\end{equation}
and we have obtained the asymptotic form of the even coefficients
\begin{equation}
\beta_{2j} = \frac{\pi}{2}\frac{1}{a^{2j}} \frac{(2^{j-1}
(j-1)!)^2}{(2j-1)!} + O\left(\frac{1}{a^{2j-1}}\right).
\end{equation}
We are now able to find the asymptotic expansion of the flux
profile (\ref{eq:flux-integral-phi}),
\begin{eqnarray*}
f(\theta) & = & -\frac{1}{2} + \frac{1}{2\pi}\frac{d}{d\theta}
\left[ \frac{\cos\ds{\frac{\theta}{2}}}{b} \left(\pi a_0 - 4
\int_0^{\sqrt{1-a^2}} \frac{\arccos \sqrt{s^2+a^2}\,
s^2\,ds}{\sqrt{a^2+s^2}
\,\left(s^2+b^2 \right)}\right) \right] \nonumber \\
&&\\
& = & -\frac{1}{2} - \frac{2}{\pi}\frac{d}{d\theta} \left[
\cos\ds{\frac{\theta}{2}} \sum_{n=0}^\infty \beta_{n+1} b^n
\right].
\end{eqnarray*}
Setting $\varepsilon\alpha=\pi-\theta$, we obtain after some
manipulations that to leading order in small $\varepsilon$ the
flux is given in the interval $-1<\alpha<1$ by
\begin{eqnarray}
\label{eq:flux-boundary-expansion} f(\alpha) & = &
-\frac{\alpha^2}{\varepsilon \sqrt{1-\alpha^2}} -
\frac{1}{\varepsilon} \sum_{n=0}^\infty
\left[\frac{\left(2^{n+1}(n+1)!\right)^2}{(2n+2)!}\alpha^2-\frac{(2^n
n!)^2}{(2n+1)!} \right](1 - \alpha^2)^{n+1/2} \nonumber \\
&&\nonumber\\
& & - \frac{\pi}{2\varepsilon}\sum_{n=0}^\infty
\left[\frac{(2n)!}{(2^n n!)^2
}-\frac{(2n+2)!(2n+2)}{(2^{n+1}(n+1)!)^2}\alpha^2
\right](1-\alpha^2)^n + O(1).
\end{eqnarray}

\noindent {\bf Acknowledgment:} This research was partially
supported by research grants from the Israel Science Foundation,
US-Israel Binational Science Foundation, and the NIH Grant No.
UPSHS 5 RO1 GM 067241.

\begin{figure}
\includegraphics{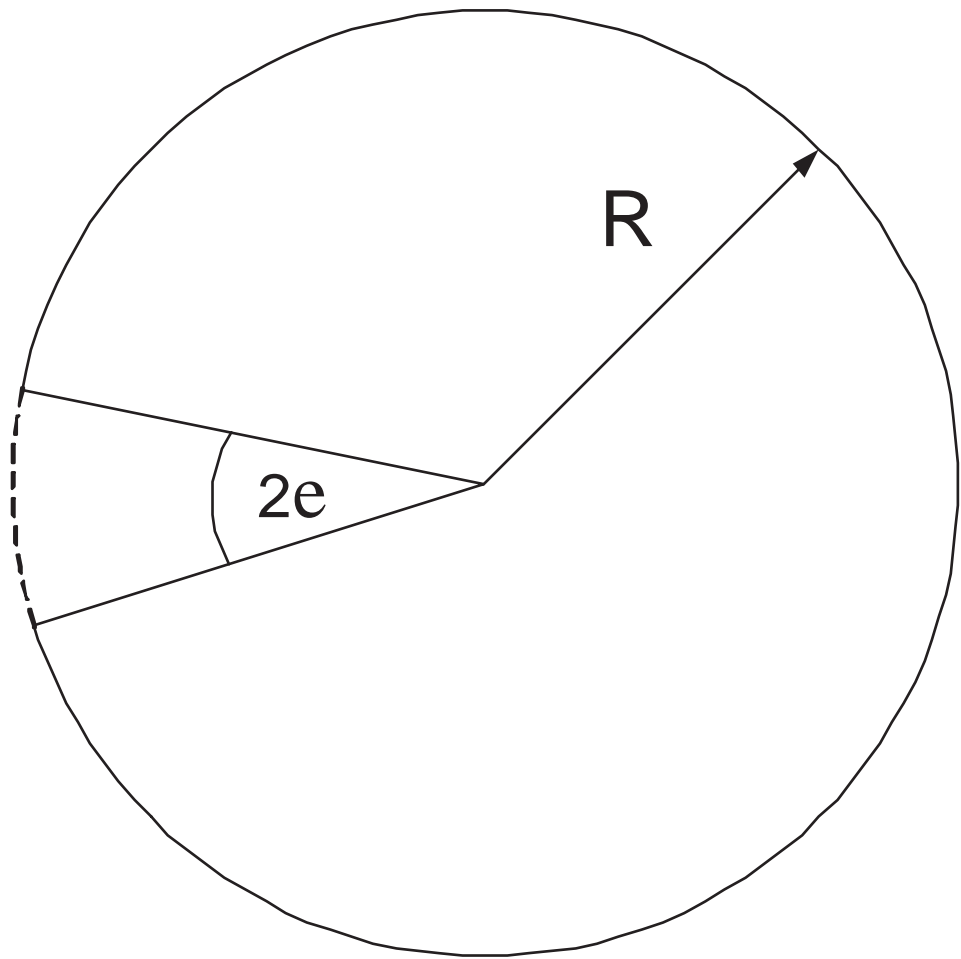}
\caption{A circular disk of radius $R$. The arclength of the
absorbing boundary (dashed line) is $2\varepsilon R$. The solid
line indicates the reflecting boundary.}\label{f:disk}
\end{figure}

\begin{figure}
\includegraphics{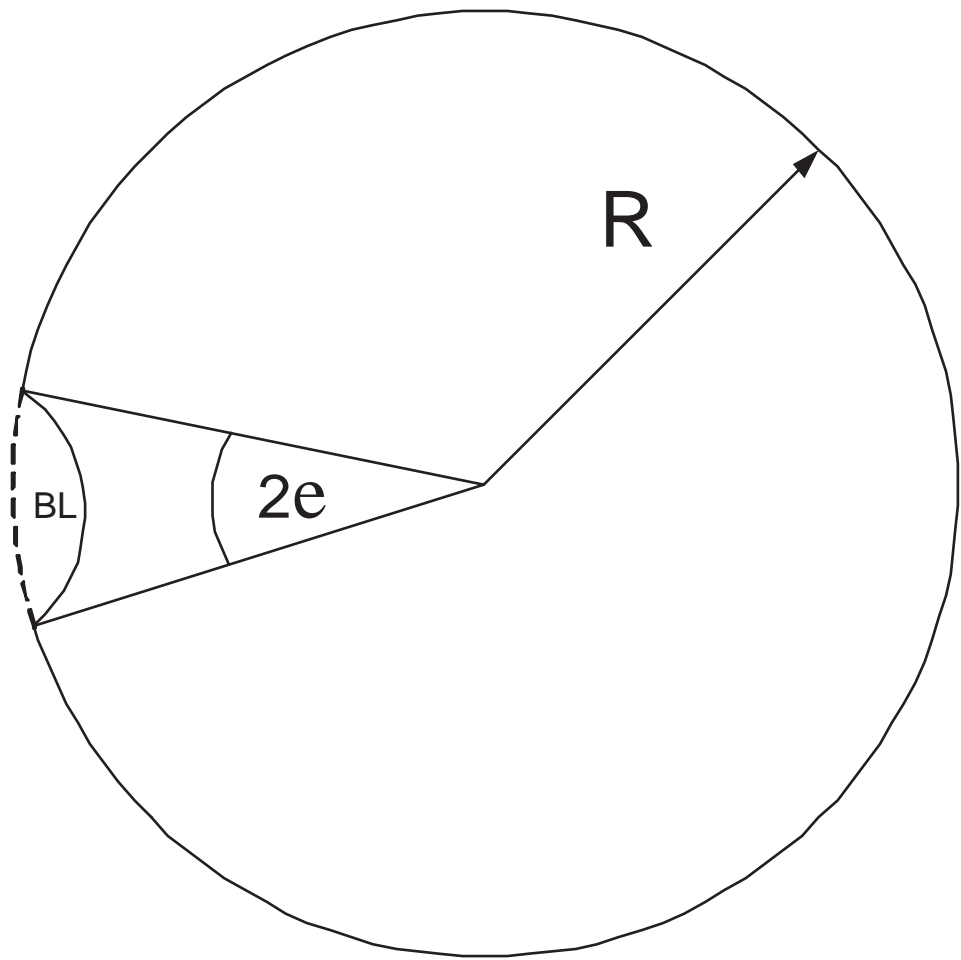}
\caption{The boundary layer, indicated by ``BL", is the area
bounded by the absorbing boundary (dashed line) and the solid arc.
Outside the boundary layer the MFPT is $O\left(\log
\ds{\frac{1}{\varepsilon}}\right)$.}\label{f:boundary-layer}
\end{figure}

\end{document}